\documentclass[11pt,a4paper]{article}
\usepackage{amsmath}
\usepackage{amssymb}
\usepackage{mathrsfs}
\usepackage{natbib}
\usepackage{graphicx}
\usepackage{hyperref}
\hypersetup{
    colorlinks=true,  
    linkcolor=blue,   
    citecolor=blue
}
\usepackage{xcolor}
\usepackage{a4wide}
\usepackage{tikz}
\usepackage{algorithm}
\usepackage{algorithmic}

\input{preamble.tex}

\reversemarginpar

\begin{document}

\title{Robust Product Markovian Quantization}

\author{Ralph Rudd\thanks{The African Institute of Financial Markets and Risk Management, University of Cape Town}
\and Thomas A. McWalter\footnotemark[1]\phantom{*}\thanks{Department of Statistics, University of Johannesburg}
\and J\"{o}rg Kienitz\footnotemark[1]\phantom{*}\thanks{Fachbereich Mathematik und Naturwissenschaften, Bergische Universit\"{a}t Wuppertal}
\and Eckhard Platen\footnotemark[1]\phantom{*}\thanks{Finance Discipline Group and School of Mathematical and Physical Sciences, University of Technology Sydney}}

\date{June 24, 2020}

\maketitle

\begin{abstract}
Recursive marginal quantization (RMQ) allows the construction of optimal discrete grids for approximating solutions to stochastic differential equations in $d$-dimensions. Product Markovian quantization (PMQ) reduces this problem to $d$ one-dimensional quantization problems by recursively constructing product quantizers, as opposed to a truly optimal quantizer. However, the standard Newton-Raphson method used in the PMQ algorithm suffers from numerical instabilities, inhibiting widespread adoption, especially for use in calibration. By directly specifying the random variable to be quantized at each time step, we show that PMQ, and RMQ in one dimension, can be expressed as standard vector quantization. This reformulation allows the application of the accelerated Lloyd's algorithm in an adaptive and robust procedure. Furthermore, in the case of stochastic volatility models, we extend the PMQ algorithm by using higher-order updates for the volatility or variance process. We illustrate the technique for European options, using the Heston model, and more exotic products, using the SABR model.
\end{abstract}



%

\section{Introduction}
Quantization is a compression technique used to approximate a given signal using less information than the original, by minimizing a measure of error called the distortion. In mathematical finance, it is used to approximate probability distributions and has been applied to the pricing of options with path dependence and early exercise \citep{PagesWilbertz2012, sagna2010pricing, bormetti2017backward}, stochastic control problems \citep{PagesPhamPrintems2004}, and non-linear filtering \citep{PagesPham2005}.

\cite{pages2015recursive} introduced a technique known as recursive marginal quantization (RMQ), which approximates the marginal distribution of a system of stochastic differential equations by recursively quantizing the Euler updates of the processes. In one dimension, the RMQ algorithm has been extended to higher-order schemes \citep{mcwalter2018}, and has been used to calibrate a local volatility model \citep{Callegaroetal2015a}.

Applying RMQ to multidimensional SDEs requires the use of stochastic numerical methods, such as the randomized Lloyd's method or stochastic gradient descent methods, e.g., Competitive Learning Vector Quantization (see \cite{pagesintroduction} for an overview of these methods). The computational cost of these techniques can be prohibitive.

\cite{callegaro2016pricing} overcame the need for stochastic methods by using conditioning to derive a modified RMQ algorithm in the context of stochastic volatility models. Their approach was to perform a standard one-dimensional RMQ on the volatility process, and then condition on the realizations of the resulting quantizer when quantizing the asset process. The modified RMQ algorithm used for the asset process retained a Newton-Raphson iteration. In doing so, they relied on the approach proposed in a preprint of \cite{fiorin2018product}.\footnote{The preprint characterized the Newton-Raphson iteration in terms of expectations of a conditioning variable, $\zeta$, see Proposition~3.5 in \url{https://arxiv.org/pdf/1511.01758v2}.} Later, in the two-dimensional case, \cite{rudd2017quantization} formulated a product quantization algorithm without the need for this conditioning, thereby increasing computational efficiency. Independently and at around the same time, an updated preprint of \cite{fiorin2018product} also removed the conditioning.\footnote{See Remark~3.4 in \url{https://arxiv.org/pdf/1511.01758v3} for the reformulated Newton-Raphson iteration.} This approach has been called product Markovian quantization (PMQ).

The contribution of the present work is two-fold. Firstly, by directly specifying the random variable to be quantized we show how both the RMQ and PMQ algorithms can be formulated as standard vector quantization. This allows us to extend the work of \cite{bormetti2017backward} and apply the accelerated Lloyd's algorithm to PMQ, should the more efficient Newton-Raphson method become unstable. Secondly, we show how to extend the higher-order quantization technique from \cite{mcwalter2018} so that it can be applied to stochastic volatility models. We now provide an overview of the paper.


In section~\ref{Sec: VQ}, the underlying mathematics of vector quantization is reviewed along with the two numerical methods central to the paper: Lloyd's algorithm and the Newton-Raphson method. For the one-dimensional case, these algorithms are specified in terms of the density function, distribution function and first lower partial expectation of the random variable being quantized. In section~\ref{Sec: RMQ}, we review the RMQ algorithm and show how it is amenable to the standard techniques of vector quantization. Section~\ref{Sec: PMQ} follows along similar lines with regards to the PMQ algorithm, and section~\ref{Sec: SV} shows how higher-order discretization schemes can be incorporated when the algorithm is applied to stochastic volatility models. Numerical results for the popular Heston and SABR models are presented in section~\ref{Sec: Numerical Results}, including exotic option pricing and a proof-of-concept calibration. Section~\ref{Sec: Conclusion} concludes the paper.

\section{Quantizing random vectors}
\label{Sec: VQ}
Let $\bX$ be a continuous random vector, taking values in $\R^d$, and defined on the probability space $\triple$. We seek an approximation of this random vector, denoted $\bXq$, taking values in a set of finite cardinality, $\Gamma$, with the minimum average squared Euclidean difference from the original. Constructing this approximation is known as \emph{quantization}, with $\bXq$ called the \emph{quantized} version of $\bX$ and the set $\Gamma = \{\bx^1, \dots, \bx^{N}\}$ known as the \emph{quantizer}, with cardinality $N$. The elements of $\Gamma$ are called \emph{codewords} or \emph{elementary quantizers}. The probabilities associated with each codeword are denoted $\P(\bXq=\bx^i)$, for $1\leq i \leq N$, and are also known as \emph{weights}.

The primary utility of quantization is the efficient approximation of expectations, or conditional expectations, of functionals of the random variable $\bX$, e.g.,
\[
    \E{H(\bX)}=\int_{\R^d}H(\bx)\,d\P(\bX\leq \bx)\approx\sum_{i=1}^N H(\bx^i)\P\big(\bXq=\bx^i\big). 
\]

We now briefly describe the mathematics of quantization. Consider the nearest-neighbor projection operator, $\pi_{\Gamma}:\R^d\mapsto\Gamma$, given by
\begin{align*}
    \pi_{\Gamma}(\bX)\vcentcolon=\big\{\bx^{i}\in\Gamma:\,\,\norm{\bX-\bx^{i}}\leq \norm{\bX-\bx^{j}}\text{ for } j=1,\dots,N,\ j\neq i\}.
\end{align*}
The quantized version of $\bX$ is defined in terms of this projection operator as $\bXq \coloneqq \pi_{\Gamma}(\bX)$. The \emph{region} $R^i(\Gamma)$, for $1\leq i \leq N$, is defined as
\[
    R^i(\Gamma)\coloneqq\big\{\mathbf{x}\in\R^d:\,\,\pi_{\Gamma}(\mathbf{x})=\bx^{i}\big\},
\]
and is the subset of $\R^d$ mapped to codeword $\bx^i$ through the projection operator. It allows the probabilities associated with each codeword to be determined as $\P(\bXq=\bx^i) =\P(\bX\in R^i(\Gamma))$.

To obtain the optimal quantizer, we must minimize the expected squared Euclidean error, known as the \emph{distortion}, given by
\begin{align*}
    D(\Gamma)&=\E[\big]{\norm{\bX - \bXq}^2}\notag\\
    &=\int_{\R^d}\norm{\bx-\pi_{\Gamma}(\bx)}^2\,d\P(\bX\leq\bx)\notag\\
    &=\sum_{i=1}^N\int_{R^i(\Gamma)}\norm{\bx-\bx^{i}}^2\,d\P(\bX\leq \bx).
\end{align*}
The symbol $\bx$ refers to the continuous domain of the distribution of the random vector $\bX$, whereas $\bx^i$ refers to the discrete codewords of the resulting quantizer, $\Gamma$, for $1\leq i \leq N$.

\subsection{Constructing optimal quantizers}
\label{Sec: Construction Optimal Quantizers}
A common fixed-point algorithm for obtaining an optimal quantization grid is known as \emph{Lloyd's algorithm} \citep{lloyd1982least}, and is based on recursively enforcing the \emph{self-consistency}\footnote{The self-consistency property is often known as \emph{stationarity} in the literature. This term is avoided to prevent potential confusion with stationary stochastic processes.} property of optimal quantizers.

By setting the gradient of the distortion to zero, it can be shown that any quantizer that minimizes the distortion function must be self-consistent, i.e.,
\[
    \bXq = \E[\big]{\bX \big|\bXq}, \qquad\text{or equivalently}\qquad \bx^{i} = \frac{\E*{\bX\ind{\bX\in R^i(\Gamma)}}}{\P(\bX\in R^i(\Gamma))},
\]
for $i = 1,\dots, N$. For an optimal quantization grid, the self-consistency condition requires that each codeword is the \emph{probability mass centroid} of its associated region.

Lloyd's algorithm iteratively enforces this condition until a desired tolerance is achieved, or the maximum number of iterations is attained, using
\begin{equation}
    \label{Eq: Lloyd d-D}
    \prescript{(l+1)}{}{\bx}^{i} = \frac{\E*{\bX\ind{\bX\in R^i\left(\prescript{(l)}{}{\Gamma}\right)}}}{\E*{\ind{\bX\in R^i\left(\prescript{(l)}{}{\Gamma}\right)}}},
\end{equation}
where $0\leq l<l_{\mathrm{max}}^{\mathrm{LA}}$ is the iteration index. In a multidimensional setting, Monte Carlo (or quasi-Monte Carlo) methods are used to compute the required expectations. 

The special case when $X$ is a one-dimensional random variable with a well-defined density function is relevant for many applications, including the recursive marginal quantization and product Markovian quantization algorithms presented later.

In one dimension, the regions associated with a quantizer may be defined directly as $R^i=\{x\in\R:\,x^{i-}<x\leq x^{i+}\}$ with
\begin{equation}
    x^{i-}\coloneqq\frac{x^{i-1}+x^{i}}{2}\qquad\text{and}\qquad x^{i+}\coloneqq\frac{x^{i}+x^{i+1}}{2}, \label{Eq: Plus-Minus Notation}
\end{equation}
for $1\leq i\leq N$, where, by definition, $x^{1-}\vcentcolon=-\infty$ and $x^{N+}\vcentcolon=\infty$. If the distribution under consideration is not defined over the whole real line, then $x^{1-}$ and $x^{N+}$ are adjusted to reflect the support. 

Suppose $f_X$ and $F_X$ are the density and distribution functions of $X$, respectively. Define the $p$-th lower partial expectation as
\[
    M_X^p(x)\vcentcolon=\E{X^p\ind{X<x}},
\]
where $M_X^0(x)=F_X(x)$ represents the distribution function of $X$. Then, direct integration of the distortion function gives
\begin{align*}
    D(\Gamma) &=\sum_{i=1}^N\int_{x^{i-}}^{x^{i+}} \norm{x-x^{i}}^2f_X(x)\,d x\\
    &=\sum_{i=1}^N\Big[M_X^2(x^{i+})-M_X^2(x^{i-})
        -2x^{i}\left(M_X^1(x^{i+})-M_X^1(x^{i-})\right)\\
        &\qquad\qquad\qquad+(x^{i})^2\left(F_X(x^{i+})-F_X(x^{i-})\right)\!\Big].
\end{align*}
Differentiating this expression with respect to each codeword, $x^i$, gives
\begin{equation}
    \label{Eq: Distortion Gradient}
    \frac{\partial D(\Gamma)}{\partial x^i}=2x^i\left(F_X(x^{i+})-F_X(x^{i-})\right)-2\left( M_X^1(x^{i+})-M_X^1(x^{i-})\right),
\end{equation}
for $1\leq i\leq N$. Thus, Lloyd's algorithm simplifies to
\begin{equation}
    \label{Eq: Lloyd 1-D}
    \prescript{(l+1)}{}{x}^{i} = \frac{M^1_X\left(\prescript{(l)}{}{x}^{i+}\right) - M^1_X\left(\prescript{(l)}{}{x}^{i-}\right)}{F_X\left(\prescript{(l)}{}{x}^{i+}\right) - F_X\left(\prescript{(l)}{}{x}^{i-}\right)},
\end{equation}
for $0\leq l<l_{\mathrm{max}}^{\mathrm{LA}}$. This means that Lloyd's algorithm may be implemented using the above closed-form expressions for the expectations given in \eqref{Eq: Lloyd d-D}, eliminating the need for Monte Carlo methods.

As an alternative to the one-dimensional Lloyd's algorithm, a Newton-Raphson iteration may be used to minimize the distortion function,
\begin{equation}
    \prescript{(l+1)}{}{\bGamma}=\prescript{(l)}{}{\bGamma}-\big[\nabla^2D\big(\prescript{(l)}{}{\bGamma}\big)\big]^{-1}\nabla D\big(\prescript{(l)}{}{\bGamma}\big), \qquad\text{with}\qquad \big[\prescript{(l)}{}{\bGamma}\big]_i\vcentcolon=\prescript{(l)}{}{x}^i, \label{Eq: Newton-Raphson}
\end{equation}
for $0\leq l<l_{\mathrm{max}}^{\mathrm{NR}}$. Here, $\prescript{(l)}{}{\bGamma}$ is a column vector, of length $N$, containing the codewords associated with the quantization grid, $\prescript{(l)}{}{\Gamma}$. What remains is to specify the gradient and the Hessian of the distortion function explicitly.

The elements of the gradient vector $\nabla D(\bGamma)$ are given directly by \eqref{Eq: Distortion Gradient}, and the tridiagonal Hessian matrix, $\nabla^2D(\bGamma)$, has diagonal elements given by
\begin{equation}
    \label{Eq: Hessian Diagonal}
    \frac{\partial^2D(\Gamma)}{\partial (x^i)^2}=2\left(F_X(x^{i+})-F_X(x^{i-})\right)
    +\tfrac12\left(f_X(x^{i+})(x^i-x^{i+1})
    +f_X(x^{i-})(x^{i-1}-x^i)\right),
\end{equation}
and super- and sub-diagonal elements given by
\begin{equation}
    \label{Eq: Hessian Off-diagonal}
    \frac{\partial^2D(\Gamma)}{\partial x^i\partial x^{i+1}}=\tfrac12f_X(x^{i+})(x^i-x^{i+1})\qquad\text{and}\qquad
    \frac{\partial^2D(\Gamma)}{\partial x^i\partial
    x^{i-1}}=\tfrac12f_X(x^{i-})(x^{i-1}-x^i),
\end{equation}
respectively. While the Newton-Raphson method has faster convergence than Lloyd's algorithm, it can become numerically unstable if the Hessian matrix, which must be inverted, becomes ill-conditioned. Thus, under certain circumstances, it may be best to use Lloyd's algorithm to ensure stability. We shall explore this issue further in section~\ref{Sec: A Robust Algorithm}.

\section{Recursive marginal quantization}
\label{Sec: RMQ}
Consider the continuous-time vector-valued diffusion, defined on the filtered probability space $(\Omega, \F, (\F_t)_{t\in[0,T]}, \mathbb{Q})$, specified by the SDE
\begin{equation}
    d\bX_t=\b{a}(\bX_t)\,dt+\b{B}(\bX_t)d\b{W}_t^\perp,\qquad \bX_0=\b{x}_0\in\R^d, \label{Eq: SDE System}
\end{equation}
with $\b{a}:\R^d\rightarrow\R^d$ and $\b{B}:\R^d\rightarrow\R^{d\times q}$, and where $\b{W}^\perp$ is a standard $q$-dimensional Brownian motion.

In general, it is not possible to form the approximation $\bXq_t\coloneqq\pi_{\Gamma_t}(\bX_t)$ by minimizing
\[
    D(\Gamma_t) = \E*{\norm{\bX_t - \pi_{\Gamma_t}(\bX_t)}^2},
\]
since the distribution of $\bX_t$ is usually unknown.

Instead we consider the discrete-time Euler approximation, $\bXd$, of $\bX$,
\begin{align}
    \bXd_{k+1}&=\bXd_{k}+ \b{a}(\bXd_k)\Delta t + \b{B}(\bXd_k)\sqrt{\Delta t}\b{z}_{k+1}\notag \\
    &=\vcentcolon\U(\bXd_{k},\b{z}_{k+1}), \label{Eq: Euler Update Function}
\end{align}
for $k = 0,\dots,K-1$, where $\Delta t =T/K$ and $\b{z}_{k+1}\sim\mathscr{N}(0,\b{I}_q)$, with initial value $\bXd_0=\b{x}_0$.

Since the distribution of $\bXd_1$ is known explicitly, standard vector quantization can be used to obtain the quantization grid at the first time step, $\Gamma_1$, and its associated probabilities. The distortion for successive time steps is then given by
\begin{align*}
    D(\Gamma_{k+1}) &= \E*{\norm{\bXd_{k+1} - \pi_{\Gamma_{k+1}}(\bXd_{k+1})}^2},
\end{align*}
for $k = 1,\dots,K-1$. However, the exact distribution of $\bXd_{k+1}$ is also unknown for $k>0$. So a further approximation is made: $\bXd_{k+1}$ is replaced by
\[ \bXa_{k+1} \vcentcolon= \U(\bXq_k, \b{z}_{k+1}), \]
a random vector that results from applying the Euler update function to the previously quantized $\bXq_k$. This gives rise to the Algorithm \ref{Alg: The RMQ Algorithm}.

\begin{algorithm}
\caption{Recursive Marginal Quantization}
\label{Alg: The RMQ Algorithm}
\begin{algorithmic}[1]
\STATE{Set $\bXq_0 \coloneqq \b{x}_0.$} \label{Alg: RMQ Random Variable}
\FOR{$k = 0$ to $K-1$}
    \STATE{Define $\bXa_{k+1} \coloneqq \U(\bXq_k, \b{z}_{k+1})$.}
    \STATE{Obtain $\Gamma_{k+1}$ by minimizing $\E*{\norm{\bXa_{k+1} - \pi_{\Gamma_{k+1}}(\bXa_{k+1})}^2}$.} \label{Alg: RMQ Grid Computation}
    \STATE{Set $\bXq_{k+1} = \pi_{\Gamma_{k+1}}(\bXa_{k+1})$.} \label{Alg: RMQ Probability Computation}
\ENDFOR
\end{algorithmic}
\end{algorithm}
Note that Step \ref{Alg: RMQ Grid Computation} computes the optimal quantization grid and Step \ref{Alg: RMQ Probability Computation} computes the associated weights, by insisting that $\P\bigl(\bXq_{k+1}=\bx^i_{k+1}\bigr) =\P\bigl(\bXa_{k+1}\in R^i(\Gamma_{k+1})\bigr)$ for $i=1,\dots,N_{k+1}$.

This procedure is known as \emph{recursive marginal quantization} and is due to \cite{pages2015recursive}. It is the repeated vector quantization of the random vector $\bXa_{k+1}$, for $k = 0,\dots,K-1$, which has distribution function
\begin{equation}
    F_{\bXa_{k+1}}(\b{x}) = \sum_{i=1}^{N_k} \Phi_{d}\left(\b{x}; \bx_k^i + \b{a}(\bx_k^i)\Delta t, \b{B}(\bx_k^i)\b{B}(\bx_k^i)^\top \Delta t \right)\P\bigl(\bXq_{k}=\bx^i_{k}\bigr), \label{Eq: Distribution Function}
\end{equation}
where $\Phi_d(\cdot; \boldsymbol{\mu}, \b{\Sigma} )$ is the $d$-dimensional Gaussian distribution function with mean $\boldsymbol{\mu}$ and covariance $\b{\Sigma}$.

In the special case where $X$ is a scalar-valued diffusion, this reduces to 
\begin{align}
    F_{\Xa_{k+1}}(x) &= \sum_{i=1}^{N_k} \Phi\left(\frac{x - c^i_k}{m^i_k} \right)p^i_k,\label{Eq: RMQ1d F}\\
    f_{\Xa_{k+1}}(x) &= \sum_{i=1}^{N_k} \frac{1}{m^i_k}\phi\left(\frac{x - c^i_k}{m^i_k} \right)p^i_k \label{Eq: RMQ1d f}\\
    \intertext{and}
    M^1_{\Xa_{k+1}}(x) &=  \sum_{i=1}^{N_k} \left[-m^i_k\phi\left(\frac{x - c^i_k}{m^i_k} \right) + c^i_k \Phi\left(\frac{x - c^i_k}{m^i_k} \right) \right]p^i_k, \label{Eq: RMQ1d M1}
\end{align}
where $\Phi(\cdot)$ and $\phi(\cdot)$ are the standard normal distribution and density functions, respectively, and
\[
    c_k^i=x_k^i+a(x_k^i),\qquad m_k^i=B(x_k^i)\sqrt{\Delta t}\qquad\text{and}\qquad p_k^i=\mathbb{P}(\Xq_{k}=x_k^i).
\]
Here, the density and lower partial expectation of $\Xa_{k+1}$ are computed by differentiating and integrating $F_{\Xa_{k+1}}(x)$, respectively. Consequently, one may now use these expressions with the standard (one-dimensional) vector quantization approaches in section~\ref{Sec: Construction Optimal Quantizers}.

Note that when using the Newton-Raphson method, \eqref{Eq: Newton-Raphson}, this is equivalent to the original approach of \cite{pages2015recursive}, with no difference in convergence characteristics. However, explicitly computing the distribution of $\Xa_{k+1}$ has the advantage that any vector quantization optimization technique may now be applied. In particular, Lloyd's algorithm, as given by \eqref{Eq: Lloyd 1-D}, may be used---this is especially useful in cases where the Newton-Raphson method fails due to numerical instability.

Furthermore, we are not limited to the Euler-Maruyama discretization, but may also use the Milstein or simplified weak-order 2.0 schemes \citep{mcwalter2018}. The necessary expressions for the latter appear in \ref{App: RMQ Marginal Distributions}.

\section{Product Markovian quantization}
\label{Sec: PMQ}
To ease the exposition, we limit ourselves to the case when $q=d$, i.e., there is one underlying Brownian motion for each dimension. Then \eqref{Eq: SDE System} can be written as
\begin{equation}
    d\bX_t=\b{a}(\bX_t)\,dt+\mathrm{diag}\left(\b{b}(\bX_t)\right)d\b{W}_t,\qquad \bX_0=\b{x}_0\in\R^d, \label{Eq: PMQ SDE}
\end{equation}
with $\b{a}(\bX_t)=[a^1(\bX_t),\dots,{a}^d(\bX_t)]^\top$, $\b{b}(\bX_t)=[b^1(\bX_t),\dots,{b}^d(\bX_t)]^\top$ and $\b{W}$ a vector of correlated Brownian motions. The correlation matrix for the Brownian motions is given by $\b{LL}^\top$, where $\b{B}$ and $\b{b}$ are related by $\b{B}(\bX_t)=\mathrm{diag}\left(\b{b}(\bX_t)\right)\b{L}$.

The marginal Euler update for each dimension is given by
\begin{align}
\Xd_{k+1}^{\nsup} &= \Xd_k^{\nsup} + a^{\nsup}(\bXd_k)\Delta t +  b^{\nsup}(\bXd_k)\sqrt{\Delta t}z^{\nsup}_{k+1} \notag\\
&=\vcentcolon\U^{\nsup}(\bXd_{k},z^{\nsup}_{k+1}), \label{Eq: Marginal Euler Update}
\end{align}
with $k = 0,\dots,K-1$. Here $\Delta t =T/K$, $z^{\nsup}_{k+1}$ is a correctly correlated Gaussian random variate, and $\Xd_0^{\nsup} = [\b{x}_0]_n$. The central idea of \emph{product Markovian quantization} (PMQ) is to quantize each of the $d$ dimensions separately, using the marginal Euler updates, and construct the required $d$-dimensional quantizer from their Cartesian product. This yields Algorithm \ref{Alg: The PMQ Algorithm}

\begin{algorithm}
\caption{Product Markovian Quantization}
\label{Alg: The PMQ Algorithm}
\begin{algorithmic}[1]
\STATE{Set $\bXq_0 \coloneqq \b{x}_0.$}
\FOR{$k = 0$ to $K-1$}
    \STATE{Define $\bXa_{k+1} \coloneqq \U(\bXq_k, \b{z}_{k+1})$.} \label{Alg: PMQ Random Variable}
    \FOR{$n=1$ to $d$}
        \STATE{Define $\Xa^\nsup_{k+1} \coloneqq \U^\nsup(\bXq_k, {z}^\nsup_{k+1})$.}
        \STATE{Obtain $\Gamma^\nsup_{k+1}$ by minimizing $\E*{\norm{\Xa^\nsup_{k+1} - \pi_{\Gamma^\nsup_{k+1}}(\Xa^\nsup_{k+1})}^2}$.}
    \ENDFOR
    \STATE{Set $\Gamma_{k+1} = \Gamma^{\gensup{1}}_{k+1}\times\dots\times\Gamma^{\gensup{d}}_{k+1}$.} \label{Alg: PMQ Grid Computation}
    \STATE{Set $\bXq_{k+1} = \pi_{\Gamma_{k+1}}(\bXa_{k+1})$.} \label{Alg: PMQ Probability Computation}
\ENDFOR
\end{algorithmic}
\end{algorithm}
Note that Algorithms \ref{Alg: The RMQ Algorithm} and \ref{Alg: The PMQ Algorithm} quantize the same random variable at each step, but differ in how they compute the underlying quantization grid. The weights associated with the computed grid are determined the same way (Steps \ref{Alg: RMQ Probability Computation} and \ref{Alg: PMQ Probability Computation}, respectively), using \eqref{Eq: Distribution Function}.

Algorithm  \ref{Alg: The PMQ Algorithm} solves $d$ one-dimensional vector quantization problems at each time step, instead of solving one $d$-dimensional quantization problem. \cite{fiorin2018product} derive a theoretical error bound for the approach.

As opposed to deriving the gradient and Hessian required for the Newton-Raphson iteration directly (the approach taken by \cite{fiorin2018product}), we derive the distribution of each $\Xa_{k+1}^{\nsup}$. We also explicitly show how to compute the joint probabilities associated with the product grid, $\Gamma_{k+1}$, for $k = 0,\dots,K-1$.

The grid at each time step is defined, in terms of the one-dimensional quantizers, as the Cartesian product $\Gamma_k=\Gamma^1_k\times\cdots\times\Gamma^d_{k}$, with cardinality $N_k=\prod_{n=1}^dN^{\nsup}_k$. Let the quantizer be specified as the enumerated set\footnote{The order of enumeration is irrelevant, as long as it is consistently applied.} of unique (product) codewords $\Gamma_k =\{ \bx^1_k,\dots, \bx^{N_k}_k \}$, with
\[ 
    \bx^i_k = \big[x^{1, i_1}_k,\dots, x^{d, i_d}_k \big]^\top,
\]
where there is a unique set of sub-indices $\{i_1,\ldots,i_d\}$, associated with index $i$. These indices enumerate the constituents of $\bx^i_k$, in terms of the underlying one-dimensional quantizers, $x^{j, i_j}_k\vcentcolon=x_k^{i_j}\in\Gamma_k^j$.
Thus, each scalar constituent of the product codeword has two superscripts: the first is its associated dimension, and the second is the index of the codeword in its corresponding one-dimensional quantizer.

With this notation in place, consider the following shorthand for the marginal Euler update from \eqref{Eq: Marginal Euler Update},
\begin{align*}
    U^{\nsup,\ivec}_{k+1} \vcentcolon= m_k^{\nsup,\ivec}z^{\nsup}_{k+1} + c_k^{\nsup,\ivec}
\end{align*}
where
\begin{align*}
    m_k^{\nsup,\ivec} = b^{\nsup}\left(\bxq_k^{\ivec}\right)\sqrt{\Delta t} \quad \text{and} \quad c_k^{\nsup,\ivec} = x_k^{n, i_n} + a^{\nsup}\left(\bxq_k^{\ivec}\right)\Delta t.
\end{align*}

\noindent Now, for each $n =1,\dots,d$,
\begin{align}
    F_{\Xa_{k+1}^{\nsup}}(x) &= \Prob{\Xa_{k+1}^{\nsup}\leq x} \notag \\
    &=\sum_{i=1}^{N_k}\cProb*{\U^{\nsup}(\bXq_k, z^{\nsup}_{k+1})\leq x}{\bXq_k=\bxq_k^{\ivec}}p_k^{\ivec} \notag
    \\
    &=\sum_{i=1}^{N_k}\Prob*{U_{k+1}^{\nsup,\ivec}\leq x}p_k^{\ivec} \notag\\
    &= \sum_{i=1}^{N_k} \Phi\left(\frac{x - c_k^{\nsup,\ivec}}{m_k^{\nsup,\ivec}}\right)p_k^{\ivec}, \label{Eq: PMQ F}
\end{align}
with $p_k^{\ivec}\vcentcolon=\Prob*{\bXq_k=\bxq_k^{\ivec}}$ the joint probability, computed at the previous time step, and $k=0,\dots,K-1$. Simple differentiation and integration then yields
\begin{align}
    f_{\Xa_{k+1}^{\nsup}}(x) &= \sum_{i=1}^{N_k} \frac{1}{m_k^{\nsup,\ivec}} \phi\left(\frac{x - c_k^{\nsup,\ivec}}{m_k^{\nsup,\ivec}}\right)p_k^{\ivec}
\intertext{and}
    M_{\Xa_{k+1}^{\nsup}}^1(x) &= \sum_{i=1}^{N_k}\left[-m_k^{\nsup,\ivec}  \phi\left(\frac{x - c_k^{\nsup,\ivec}}{m_k^{\nsup,\ivec}}\right) +  c_k^{\nsup,\ivec}  \Phi\left(\frac{x - c_k^{\nsup,\ivec}}{m_k^{\nsup,\ivec}}\right)  \right] p_k^{\ivec}. \label{Eq: PMQ M1}
\end{align}
Comparing the above three equations with \eqref{Eq: RMQ1d F} to \eqref{Eq: RMQ1d M1}, the only difference is the extra superscript index, $n$, in the expressions indicating the dimension being quantized. As before, each marginal quantization in the PMQ algorithm only relies on weighted summations of the standard Gaussian density and distribution functions. Once again, using \eqref{Eq: PMQ F} to \eqref{Eq: PMQ M1} the algorithm can utilize either the one-dimensional Newton-Raphson method or the one-dimensional Lloyd's algorithm from section~\ref{Sec: Construction Optimal Quantizers}.

The probabilities associated with the final product grid, $\Gamma_{k+1}$, are computed using $F_{\bXa_{k+1}}$, the distribution of $\bXa_{k+1}$, which in turn is the joint distribution of the marginal Euler updates, $\Xa_{k+1}^{1},\dots, \Xa_{k+1}^{d}$. The regions associated with the product quantizer can only be $d$-dimensional rectangles, as they result from the Cartesian product of one-dimensional grids. Thus, the joint probability of each region can always be computed as sums and differences of multivariate normal distribution functions. For example, in the two dimensional case, we have
\begin{align*}
    p_{k+1}^i=\Prob*{\bXq_{k+1}=\bxq_{k+1}^i} &= 
    F_{\bXa_{k+1}}\left(\big[x^{1,\,i_1+}_{k+1},x^{2,\,i_2+}_{k+1}\big]^\top\right)-
    F_{\bXa_{k+1}}\left(\big[x^{1,\,i_1+}_{k+1},x^{2,\,i_2-}_{k+1}\big]^\top\right) \\
    &\quad-
    F_{\bXa_{k+1}}\left(\big[x^{1,\,i_1-}_{k+1},x^{2,\,i_2+}_{k+1}\big]^\top\right)+
    F_{\bXa_{k+1}}\left(\big[x^{1,\,i_1-}_{k+1},x^{2,\,i_2-}_{k+1}\big]^\top\right)\!,
\end{align*}
which is expressed in terms of \eqref{Eq: Distribution Function}, using the notation given in \eqref{Eq: Plus-Minus Notation}.

\subsection*{Example: Two correlated assets.}

As a first example, we consider two negatively correlated assets, each driven by geometric Brownian motion.
The SDEs for the assets may be specified in the notation of \eqref{Eq: PMQ SDE} as
\[\b{a}(\bX_t)= [rX^1_t,\, rX^2_t]^\top \qquad\text{and}\qquad \b{b}(\bX_t)= [\sigma_1 X^1_t,\,\sigma_2X^2_t]^\top, \] 
with $d\langle W^1, W^2  \rangle_t = \rho\, dt$. The parameters chosen were $\b{x}_0 = [110,\, 90]^\top$, $\sigma_1=10\%$, $\sigma_2=30\%$, $\rho=-0.6$ and $r=5\%$ for the risk-free rate. A total of $200$ codewords were used for the RMQ algorithm. For the PMQ algorithm, $N_k^1=10$ codewords were used for the marginal $X^{1}$ process, denoted Asset $1$, and $N_k^2=20$ codewords were used for the marginal $X^{2}$ process, denoted Asset $2$. Both algorithms used monthly time steps. 

\begin{figure}[t!]
    \begin{center}
        \includegraphics[width=\columnwidth]{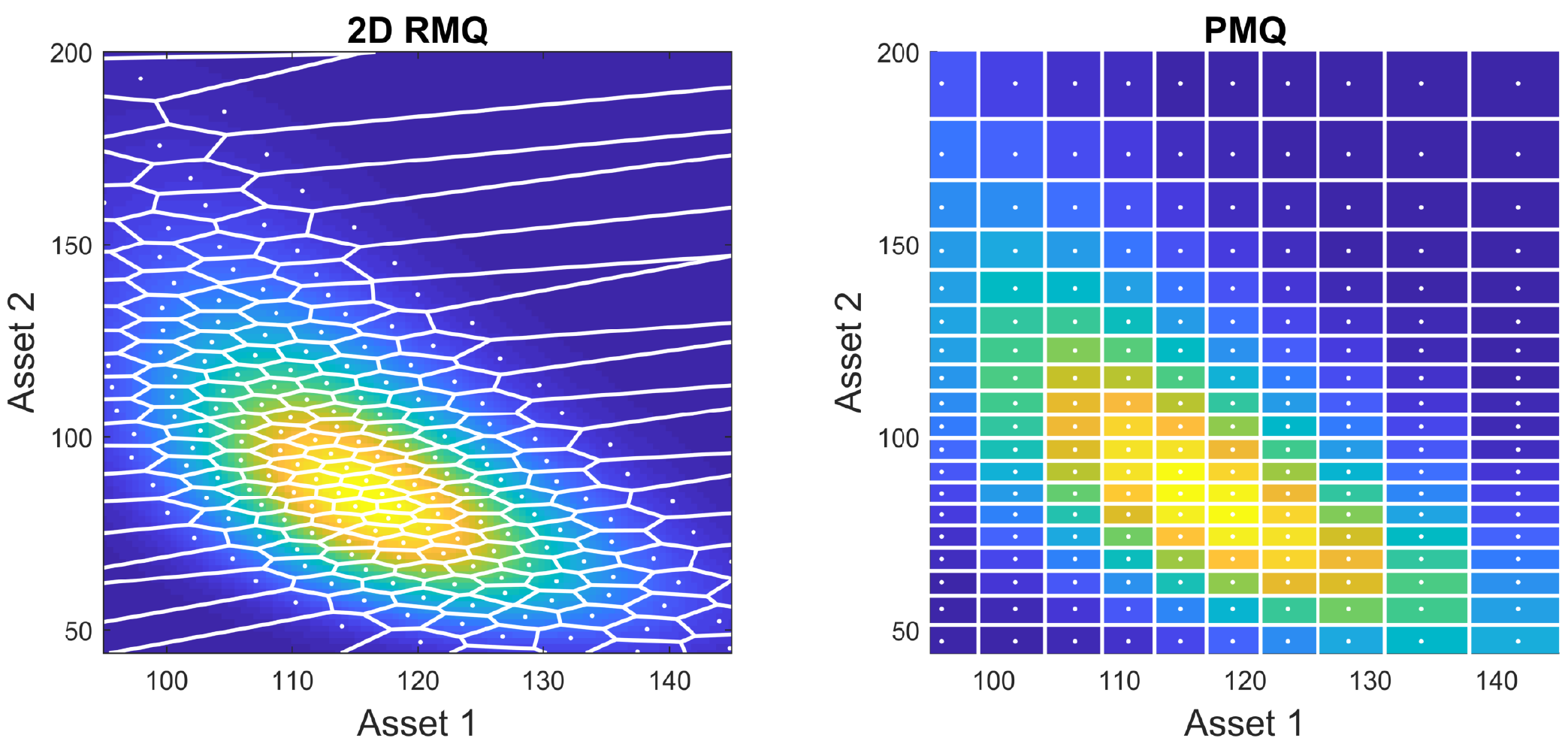}
    \end{center}
    \caption{Comparison of recursive marginal quantization and product Markovian quantization for two correlated assets.}
    \label{Fig: 2D GBM}
\end{figure}

Figure \ref{Fig: 2D GBM} illustrates the joint distribution of the assets one year into the future using the RMQ and PMQ algorithms. It highlights the fundamental differences between the two approaches. The left panel shows the codewords that result from the RMQ algorithm and their corresponding regions. The underlying heat-map represents the actual bivariate lognormal density of the two assets. The regions are polygons, with the codewords clustered in the areas of high probability. In the right panel, the heat-map represents the probability associated with each rectangular region of the product grid that results from the Cartesian product of the two marginal quantizers. Although the shape of the underlying probability density is still well approximated, the PMQ algorithm produces a higher concentration of codewords in regions of low probability, e.g., compare the upper right corner of the panels.

It is worth noting that, because of the need for stochastic methods, the RMQ algorithm is significantly slower than the PMQ algorithm, requiring approximately 200 times longer to compute.

\section{Stochastic volatility models}
\label{Sec: SV}
In the specific case of stochastic volatility models, the dependence between the asset price process and the volatility or variance process is usually less general than that allowed by \eqref{Eq: SDE System}. 
Consider a two-dimensional system of SDEs given by 
\begin{equation*}
\b{a}(\bX_t)= \left[a^1(\bX_t),\, a^2(X^2_t)\right]^\top \qquad\text{and}\qquad \b{b}(\bX_t)= \left[b^1(\bX_t),\,b^2(X^2_t)\right]^\top,
\end{equation*}
with $d\langle W^1, W^2  \rangle_t = \rho\, dt$. Note that the asset price process, $X^1$, does not appear in the drift or diffusion coefficient of the volatility or variance process, $X^2$. It should be clear that in this case, the marginal quantization of $X^{2}$ can be completed for all $k=0,\dots,K$ without reference to the $X^{1}$ process. This allows higher-order updates to be used for $X^{2}$ in the same way as for the one-dimensional RMQ algorithm \citep{mcwalter2018}. The PMQ algorithm remains unchanged, except that the joint probabilities must be computed using a new joint distribution. We illustrate the case when the simplified weak-order 2.0 scheme is used. 

To derive the required joint distribution, we adopt the short-hand notation of \ref{App: RMQ Marginal Distributions}\ for the $X^{2}$ process, modified slightly to incorporate another index in the superscript to indicate the second dimension. Let $\Phi_2(x, y, \rho)$ be the bivariate Gaussian cumulative distribution function evaluated at $[x, y]^\top$ with correlation $\rho$. The joint distribution becomes
\begin{align*}
    F_{\bXa_{k+1}}\big([x, y]^\top\big) &=\sum_{i=1}^{N_k} \cProb*{z^{1}_{k+1}\leq \frac{x - c^{1,\,i}_k}{{m}^{1,\,i}_k}, \left(z^{2}_{k+1}+ \sqrt{\bar{\lambda}^{2,\,i}_k}\right)^2\leq \frac{y - \bar{c}^{2,\,i}_k}{\bar{m}^{2,\,i}_k}}{\bXq_k = \bx^i_k}  p^{i}_k \\
    &= \sum_{i=1}^{N_k}\left[ \cProb*{z^{1}_{k+1}\leq \frac{x - c^{1,\,i}_k}{{m}^{1,\,i}_k}, z^{2}_{k+1}\leq -\sqrt{\bar{\lambda}^{2,\,i}_k} + \sqrt{\frac{y - \bar{c}^{2,\,i}_k}{\bar{m}^{2,\,i}_k}}}{\bXq_k = \bx^i_k}\right.\\
    &\qquad\left. - \cProb*{z^{1}_{k+1}\leq \frac{x - c^{1,\,i}_k}{{m}^{1,\,i}_k}, z^{2}_{k+1}\leq -\sqrt{\bar{\lambda}^{2,\,i}_k} - \sqrt{\frac{y - \bar{c}^{2,\,i}_k}{\bar{m}^{2,\,i}_k}}}{\bXq_k = \bx^i_k}\right]p^{i}_k\\
    &= \sum_{i=1}^{N_k}\left[\Phi_2\left(\frac{x - c^{1,\,i}_k}{{m}^{1,\,i}_k}, -\sqrt{\bar{\lambda}^{2,\,i}_k} + \sqrt{\frac{y - \bar{c}^{2,\,i}_k}{\bar{m}^{2,\,i}_k}}; \rho \right) \right.  \\
    &\qquad\left.- \Phi_2\left(\frac{x - c^{1,\,i}_k}{{m}^{1,\,i}_k}, -\sqrt{\bar{\lambda}^{2,\,i}_k} - \sqrt{\frac{y - \bar{c}^{2,\,i}_k}{\bar{m}^{2,\,i}_k}}; \rho \right)\right]p^{i}_k,
\end{align*}
which has the net effect of adding an additional evaluation of the bivariate normal distribution function to each term in the summation. Although currently no theoretical proof of convergence exists, the potential effectiveness of this technique is illustrated in section~\ref{Sec: Numerical Results}.

\section{A robust algorithm}
\label{Sec: A Robust Algorithm}
In the literature, the Newton-Raphson method is used extensively for the RMQ algorithm in one dimension \citep{pages2015recursive, Callegaroetal2015a, callegaro2016pricing}. \cite{fiorin2018product} states that \emph{fast} quantization is only available in one-dimension because of deterministic procedures like the Newton-Raphson method, and this motivates the derivation of the product Markovian quantization technique.

However, the Newton-Raphson method has three flaws when used to solve the one-dimensional vector quantization problems that arise in RMQ and PMQ. Firstly, should negative codewords be generated when the process dynamics excludes crossing the zero boundary, the method will fail. This can occur because we are approximating the true distribution of the process using Euler updates, which may admit negative values. \cite{mcwalter2018} solved this problem by showing how to correctly model the zero boundary to ensure positive codewords.

Secondly, the Newton-Raphson method is sensitive to the initial guess used, and may fail to converge. Thirdly, the Hessian matrix may become ill-conditioned, which may result in significant numerical error in the solution of the linear system. \cite{bormetti2017backward} briefly explore both these conditions and derive an RMQ-specific Lloyd's algorithm to address them.

By directly specifying the random variable to be quantized, we have shown how one-dimensional RMQ and $d$-dimensional PMQ can both be solved using either the standard Newton-Raphson method or the one-dimensional Lloyd's algorithm, without modification. 

In Algorithm \ref{Alg: Algorithm Figure}, we recommend a hybrid approach, where the Newton-Raphson method is applied until either the iteration limit, $l_{\text{max}}^{\mathrm{NR}}$, is attained or the Hessian matrix becomes numerically unstable, at which point we switch to Lloyd's algorithm to complete the quantization, using a maximum number of iterations $l_{\text{max}}^{\mathrm{LA}}$.

For line \ref{Alg: Lloyds 2} in Algorithm \ref{Alg: Algorithm Figure}, we define the function $g(\cdot)$ to apply \eqref{Eq: Lloyd 1-D} to each element of the $\bGamma$ vector, such that
\[ [g(\bGamma)]_i = \frac{M^1_X\left({x}^{i+}\right) - M^1_X\left({x}^{i-}\right)}{F_X\left({x}^{i+}\right) - F_X\left({x}^{i-}\right)}.\]
In this way, Lloyd's algorithm is expressed as a general fixed point algorithm and can immediately benefit from Anderson acceleration \citep{walker2011anderson}. 

This approach allows us to leverage the speed of the Newton-Raphson method, while falling back to the robust accelerated Lloyd's algorithm when required. This is essential for applications like calibration, see section~\ref{Sec: Calibration}.

The basic Anderson acceleration algorithm is described in \cite{bormetti2017backward} and a complete MATLAB implementation is provided in \cite{walker2011implementation}. Furthermore, in MATLAB, monitoring the Hessian matrix can be done at no further computational cost. In our implementation we rely on the LU-decomposition to solve the linear system in \eqref{Eq: Newton-Raphson}. By default, MATLAB will issue a warning when the matrix to be decomposed is close to singular. By escalating this warning to an error, we can use exception handling to switch between the Newton-Raphson method and the accelerated Lloyd's algorithm.

\begin{algorithm}
\caption{Calculating $\bGamma$}
\label{Alg: Algorithm Figure}
\begin{algorithmic}[1]
\STATE{Set $\bGamma = \bGamma_{k-1}$ and $l = 1$}
\WHILE{$\text{cond}(\nabla^2D\left({\bGamma}\right)<\mathit{tol}$ and $l \leq l_{\text{max}}^{\mathrm{NR}}$}
\STATE{$\bGamma \leftarrow \bGamma - \left[\nabla^2D\left({\bGamma}\right)\right]^{-1}\nabla D\left(\bGamma \right)$}
\STATE{$l \leftarrow l + 1$}
\ENDWHILE
\IF{$l < l_{\text{max}}^{\mathrm{NR}}$}
\FOR{$l=1$ to $l_{\text{max}}^{\mathrm{LA}}$} \label{Alg: Lloyds 1}
\STATE{$\bGamma \leftarrow g\left(\bGamma\right)$} \label{Alg: Lloyds 2}
\ENDFOR \label{Alg: Lloyds 3}
\ENDIF
\end{algorithmic}
\end{algorithm}

\section{Numerical results}
\label{Sec: Numerical Results}
In this section, we price options under the \cite{heston1993closed} and SABR \citep{hagan2002managing} models, and provide a proof-of-concept calibration for the SABR model. For European options, the Heston model is amenable to semi-analytical pricing using Fourier transform techniques, whereas an analytical approximation exists for both the Black and Bachelier implied volatilities under the SABR model. The Fourier pricing technique implemented uses the little trap formulation of the characteristic function for the Heston model \citep{albrecher2006little}, while the implied volatility approximation for the SABR model is the latest from \cite{hagan2016universal}.

The Heston example serves to highlight the effect of changing the discretization of the independent process. For the SABR model we price up-and-out barrier and Bermudan put options, and provide a proof-of-concept calibration to market data, illustrating the flexibility of the PMQ algorithm.

All simulations were executed using MATLAB R2018a on a computer with a $2.20$ GHz Intel i-$7$ processor and $8$ GB of RAM.




\subsection{The Heston model}
\begin{figure}
    \begin{center}
        \includegraphics[width=\columnwidth]{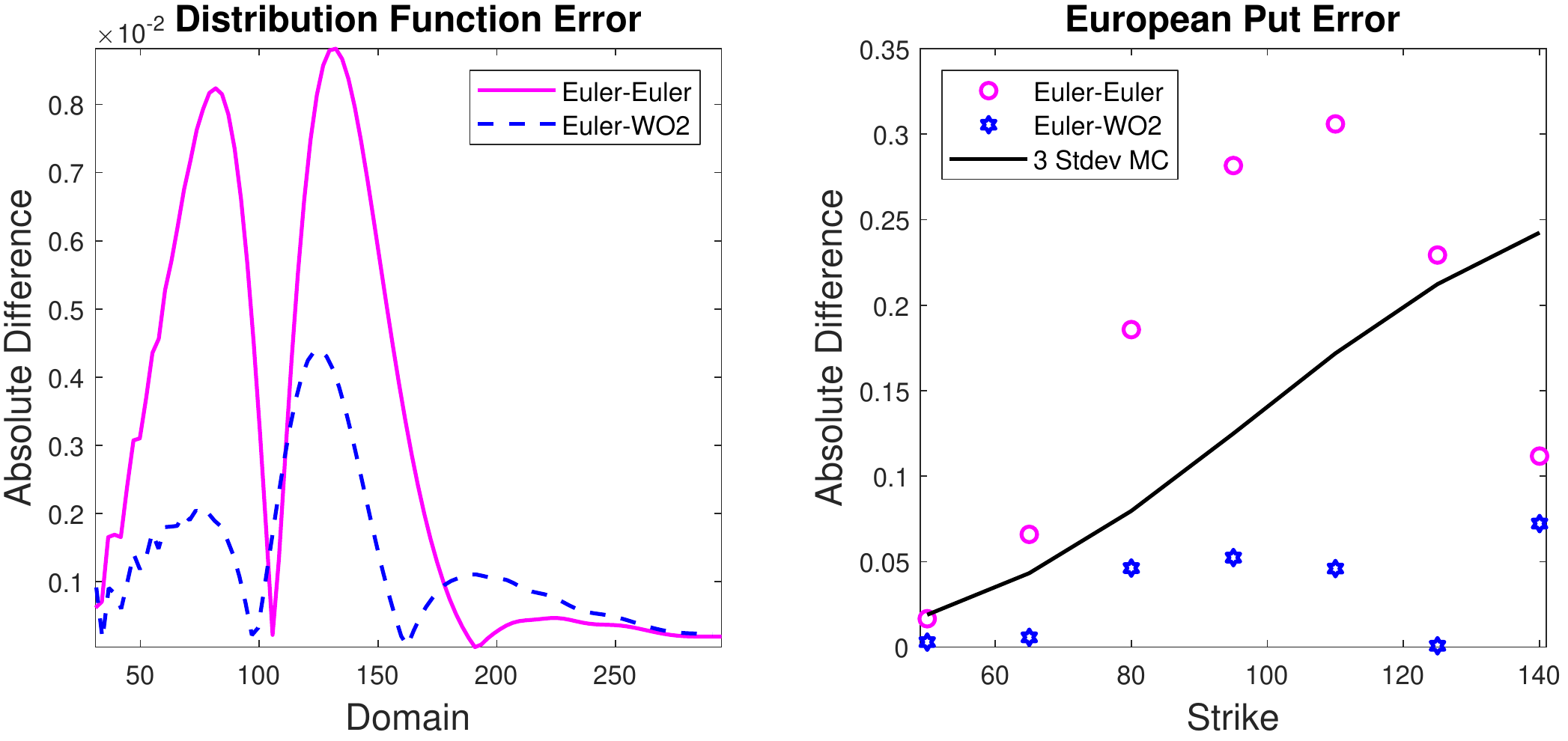}
    \end{center}
    \caption{Comparison of the Euler-Euler PMQ algorithm and Euler-WO2 PMQ algorithm for the Heston model.}
    \label{Fig: Heston}
\end{figure}

The SDEs for the Heston model may be specified as
\[\b{a}(\bX_t)= \left[rX^1_t,\, \kappa(\theta - X_t^{2})\right]^\top \qquad\text{and}\qquad \b{b}(\bX_t)= \biggl[\sqrt{ X_t^{2}}X_t^{1},\,\sigma\sqrt{X_t^{2}}\biggr]^\top, \] 
with $d\langle W^1, W^2  \rangle_t = \rho\, dt$. The parameters chosen were $\kappa = 2$, $\theta = 0.09$, $\sigma = 60\%$, $r = 5\%$, $\rho=-0.3$, $x^{1}_0=100$ and $x^{2}_0 = 0.09$, which are based on the SV-I parameter set from Table 3 of \cite{lord2010comparison}, with $\sigma$ adjusted from $1$ to $0.6$ so that the the square-root variance process lies on the Feller-boundary, i.e., $2\kappa\theta=\sigma^2$.  For the PMQ algorithm $N^{1}=30$ and $N^{2}=15$ codewords were used for the two processes, with the number of codewords held constant through time. The maturity was set at $T=1$, and $K=12$ time steps were used.

Figure \ref{Fig: Heston} demonstrates pricing a one-year European put option using the PMQ algorithm with both the standard Euler discretization for the variance process and the simplified weak-order 2.0 discretization. We denote these the Euler-Euler and Euler-WO2 schemes respectively. 

The left panel shows the absolute difference between the continuous marginal distribution of $\Xa_K$ (before quantization) and the true marginal distribution of $X_T$, which can be obtained by numerically integrating the characteristic function. As we are comparing two distribution functions, this error must be between zero and one. Although both the Euler-Euler and Euler-WO2 schemes approximate the true distribution function well, the Euler-WO2 scheme has an average absolute error across the specified domain of $0.00129$, less than half that of the Euler-Euler scheme, which has an average error of $0.00292$.

The right panel illustrates the absolute error between the price provided by quantization and the semi-analytical price for a European put over a range of strikes. It also displays the three-standard-deviation bound for a $100\,000$ path Monte Carlo simulation, which utilized the quadratic-exponential scheme of \cite{andersen2008simple}, which is neatly summarized by \cite{rouah2013heston}. It is clear that the Euler-WO2 scheme significantly outperforms the Euler-Euler scheme in terms of accuracy.

However, this may not be true in general. Improving the discretization of the volatility or variance process need not necessarily improve the grid for the asset process or the joint grid. This is because the PMQ algorithm optimizes the marginal grids separately. In this case, because we are close to violating the Feller condition for the square-root variance process, the Euler discretization is a poor choice and we can significantly improve the resulting joint grid by using the simplified weak-order 2.0 scheme. A similar situation arises in practice, as model parameters obtained by calibrating to the market often violate the Feller condition. This is also the case with the parameter set specified for the Heston model in \cite{fiorin2018product}; the prices obtained can be significantly improved by using the Euler-WO2 scheme.

\subsection{The SABR model}
\begin{figure}
    \begin{center}
        \includegraphics[width=\columnwidth]{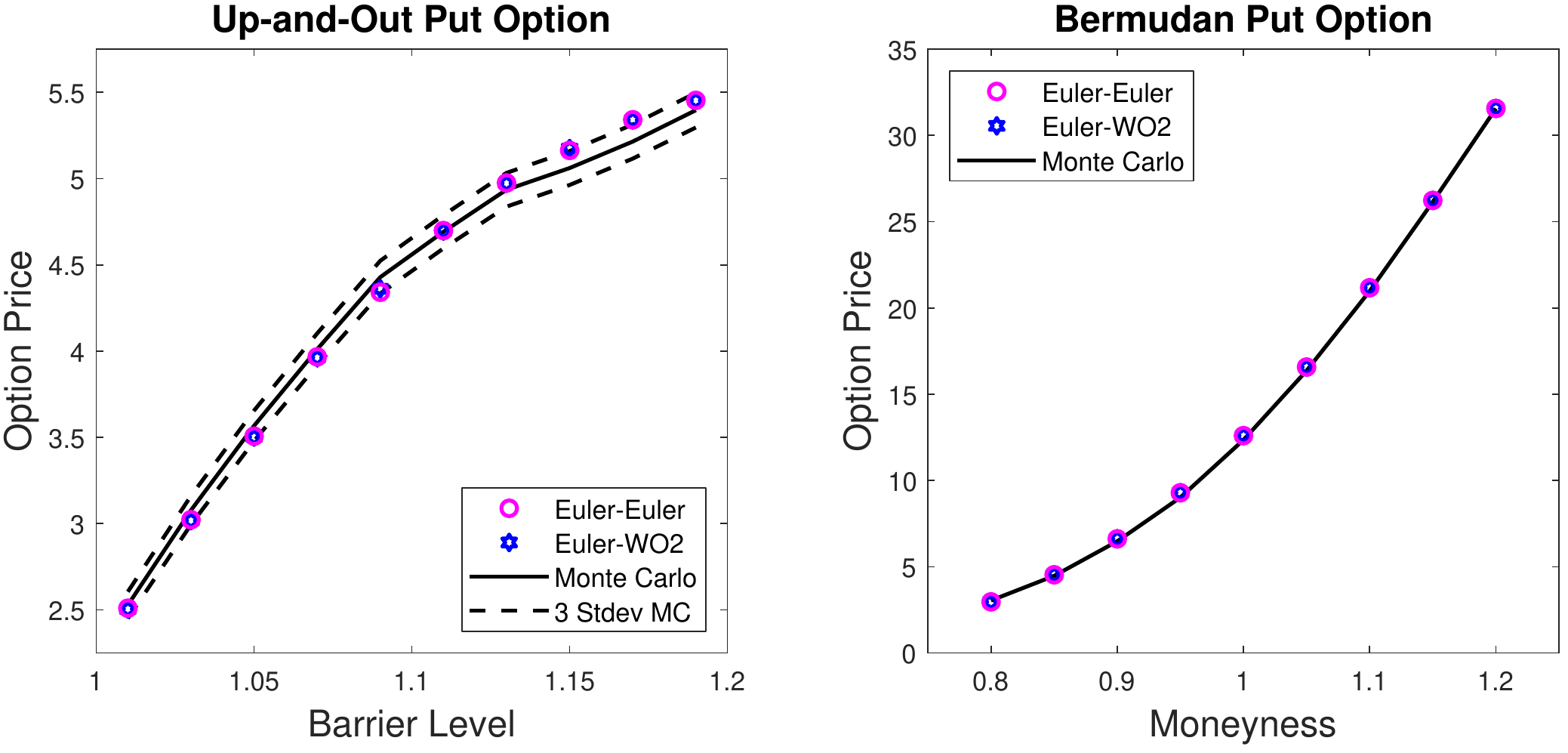}
    \end{center}
    \caption{The PMQ algorithm for pricing an up-and-out and Bermudan put options under the SABR model.}
    \label{Fig: SABR}
\end{figure}

The SDEs for the SABR model may be specified as
\[\b{a}(\bX_t)= \left[0,\, 0\right]^\top \qquad\text{and}\qquad \b{b}(\bX_t)= \left[X^2_t(X^1_t)^\beta,\, \nu X^2_t\right]^\top, \] 
with $d\langle W^1, W^2  \rangle_t = \rho\, dt$. The parameters chosen for option pricing were $\beta = 0.9$, $\nu = 0.4$, $\rho= -0.3$ and $x^{2}_0=0.4$. For this example, we model the forward value of an asset with stochastic volatility under the assumption of a constant interest rate of $r = 10\%$, such that $x^{1}_0= S_0\exp(rT)$, with $S_0 = 100$ and $T$ set at one year. For the PMQ algorithm we chose $N^1=60$, $N^2=30$ and $K = 12$. The Monte Carlo simulations in this section utilize the fully-truncated Euler scheme, suggested as the least-biased scheme for stochastic volatility models by \cite{lord2010comparison}, with $100\,000$ paths and $120$ time steps.

In the left panel of Figure \ref{Fig: SABR}, we price discrete up-and-out put options with a maturity of one year, monthly barrier evaluations and a strike of $100$, for a variety of barrier levels. The barrier levels are expressed as a percentage of strike. Two prices produced by the PMQ algorithm lie outside the three-standard-deviation bounds of the Monte Carlo simulation. However, when using the Monte Carlo prices as a benchmark, the resulting prices are very accurate, with an average relative error across the barrier levels of less than $0.1\%$.
In contrast to the Heston example, utilizing the Euler-WO2 discretization provides almost no improvement when compared to the Euler-Euler discretization. This is because there is no drift term in the volatility process and the higher-order derivatives of its diffusion term are zero, so the effect of the higher-order discretization is minimal. 

In the right panel of Figure \ref{Fig: SABR}, we price Bermudan put options, with a maturity of one year and monthly exercise opportunities, for a range of strikes. Again, if we use the high-resolution Monte Carlo simulation as a benchmark, the PMQ prices are very accurate, with an average relative error across the strikes of less than $1\%$.

\subsection{Calibration}
\label{Sec: Calibration}
\begin{figure}
    \begin{center}
        \includegraphics[width=\columnwidth]{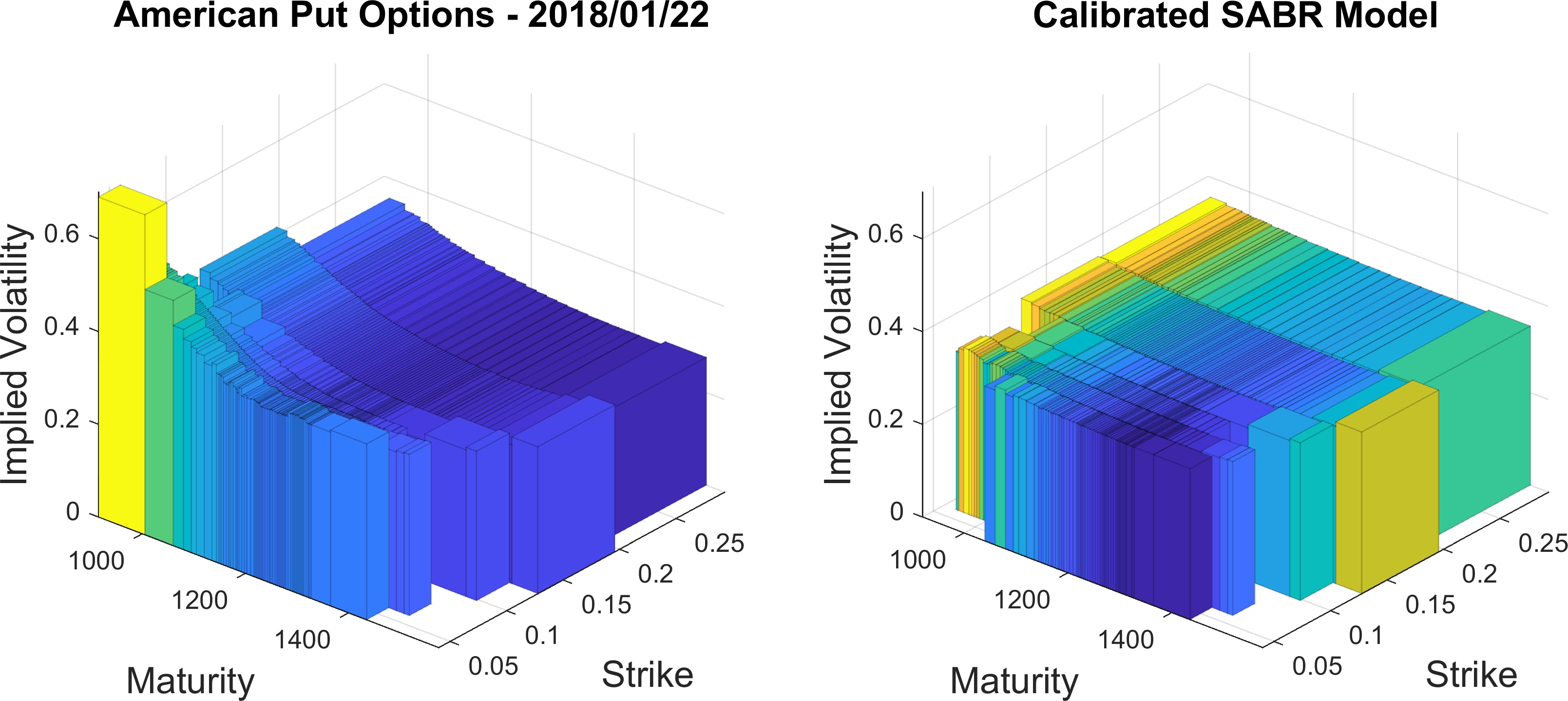}
    \end{center}
    \caption{Calibrating the SABR model to American put options on AMZN for January 22, 2018.}
    \label{Fig: Calibration}
\end{figure}

An advantage of the PMQ algorithm, like traditional tree methods, is the ability to price multiple options without needing to re-generate the underlying grid. Once the optimal quantization grid has been generated out to the furthest required option maturity, the computational cost of pricing options is negligible. An immediate application is the ability to calibrate stochastic volatility models directly to non-vanilla products. RMQ has previously been used to calibrate the quadratic normal volatility model to vanilla options on the DAX index by \cite{Callegaroetal2015a} and PMQ calibration has been demonstrated using the Heston model by \cite{Callegaro2018}.

The SABR calibration problem can be formulated as
\[\min_{\Theta \in\R^4} F(\Theta), \]
where $F$ is the objective or error function and $\Theta = \{y_0,\ \beta,\ \nu,\ \rho\}$ is the parameter set for the SABR model. 
In \cite{escobar2016parameters} the \emph{relative squared volatility error} (RSVE) is recommended as the objective function for calibrating the Heston model, and it is adopted here. 
It is defined as
\[  F(\Theta) = \sum_{l=1}^{L}\left(\frac{\sigma_l^{\mathrm{Model}}(\Theta) - \sigma_l^{\mathrm{Market}} }{\sigma_l^{\mathrm{Market}}} \right)^2, \]
where $L$ is the number of calibration instruments used, $\sigma_l^{\mathrm{Model}}(\Theta)$ is the Black-Scholes implied volatility that corresponds to pricing calibration instrument $l$ with the model parameters $\Theta$, and $\sigma_l^{\mathrm{Market}}$ is the implied volatility for that instrument observable in the market.

As a proof-of-concept example, we calibrate the SABR model directly to American put options on AMZN for January 22, 2018. We considered maturities from $3$ days to $3$ months and all strikes within $30\%$ of at-the-money that had non-zero volume, for a total of $393$ calibration instruments.
The stock price was $x^{1}_0 = 1327.31$.

The results are displayed in Figure \ref{Fig: Calibration} with the RSVE and calibrated parameters summarized in Table \ref{Tab: SABR Calibration}. Despite the poor correspondence to the extreme skew for the shortest maturity options, the calibration results in an average absolute error of less than $11\%$ across the entire volatility surface, and an average relative error of less than $2\%$.

The calibration exercise highlights the advantage of using a robust algorithm. Problems arise with the traditional Newton-Raphson method for the PMQ algorithm: at various points during the optimization, the Hessian matrix becomes ill-conditioned and fails to invert. Thus, it is necessary to switch to the one-dimensional Lloyd's algorithm as proposed in Algorithm~\ref{Alg: Algorithm Figure}.

\begin{table}[t]
    \begin{center}
        \begin{tabular}{c c c c | c }
           $x^{2}_0$ & $\beta$ &$\nu$  &$\rho$ &$F(\Theta)$\\
           $0.87$ & $0.86$ &$0.78$ &$-0.92$ &$7.36$
        \end{tabular}
    \end{center}
    \caption{Summary of calibration results for the SABR model calibrated to American put options on AMZN for January 22, 2018.}
    \label{Tab: SABR Calibration}
\end{table}

\section{Conclusion}
\label{Sec: Conclusion}
In this paper, we have formulated the one-dimensional RMQ and $d$-dimensional PMQ algorithms as standard vector quantization problems by deriving the density, distribution and lower partial expectation functions of the random variables to be quantized at each time step. As a consequence, this allows the straightforward application of Lloyd's algorithm in the cases where the traditional Newton-Raphson method becomes unstable. We proposed a hybrid algorithm that utilizes the speed of the Newton-Raphson method but may fall back to the less efficient accelerated Lloyd's algorithm when necessary.

Furthermore, we extended the PMQ algorithm for stochastic volatility models by using a simplified weak-order 2.0 update for the volatility process. The effectiveness of this technique was demonstrated by comparing the resulting marginal distributions of correlated geometric Brownian motion asset processes, and by pricing European options under the Heston model.

Finally, we priced up-and-out barrier and Bermudan put options under the SABR model and provided a proof-of-concept calibration to an American put option implied volatility surface. The calibration, in particular,  highlighted the need for our hybrid algorithm, as the inversion of the Hessian matrix became numerically unstable at various stages during the search through the parameter space.

%

\clearpage
\appendix
\renewcommand\thesection{Appendix \Alph{section}}

\section{The simplified weak-order 2.0 approximation}
\label{App: RMQ Marginal Distributions}
Consider the continuous-time scalar-valued diffusion specified by the SDE
\begin{equation}
    dX_t=a(X_t)\,dt+b(X_t)\,dW_t,\qquad X_0=x_0\in\R,\label{Eq: X SDE}
\end{equation}
defined on the filtered probability space $(\Omega, \F, (\F_t)_{t\in[0,T]}, \P)$, where $W$ is a standard one-dimensional Brownian motion.
The simplified weak-order 2.0 approximation to this process can be written as
\begin{align*}
    \Xd_{k+1} &= \bar{m}(\Xd_k) \left(z_{k+1} + \sqrt{\bar{\lambda}(\Xd_k)}\right)^2 + \bar{c}(\Xd_k), \qquad \Xd_0 = x_0,
    \intertext{with}
    \bar{m}(x)&=\tfrac{1}{2}b(x)\dx{b}{x}\Delta t,\\
    \bar{c}(x)&=x+\left(a(x)-\tfrac12b(x)\dx{b}{x}\right)\Delta t
    +\tfrac{1}{2}\left(a(x)\dx{a}{x} + \tfrac{1}{2}\ddx{a}{x}b^2(x)\right)(\Delta t)^2\\
    &\qquad-\frac{\left(b(x)+\frac{1}{2}\left(\dx{a}{x}b(x) + a(x)\dx{b}{x} + \frac{1}{2}\ddx{b}{x}b^2(x)  \right)\Delta t \right)^2}{2b(x)\dx{b}{x}},\\
    \intertext{and}
    &\bar{\lambda}(x) = \left(\frac{b(x) + \frac{1}{2}\left(\dx{a}{x}b(x) + a(x)\dx{b}{x} + \frac{1}{2}\ddx{b}{x}b^2(x) \right)\Delta t}{b(x)\dx{b}{x} \sqrt{(\Delta t)}}\right)^2.
\end{align*}
The required distribution, density and lower partial expectation functions become
\begin{align*}
    F_{\Xa_{k+1}}(x) &= \sum_{i=1}^{N_k} F_\chi \left(\frac{x - \bar{c}^i_k}{\bar{m}^{i}_k} ; 1, \lambda^i_k\right)p^{i}_k,\\
    f_{\Xa_{k+1}}(x) &= \sum_{i=1}^{N_k} \frac{1}{\bar{m}^{i}_k}f_\chi\left(\frac{x - \bar{c}^i_k}{\bar{m}^{i}_k} ;1,\lambda^i_k \right)p^{i}_k, \\
    \intertext{and}
    M^1_{\Xa_{k+1}}(x) &=  \sum_{i=1}^{N_k} \left[\bar{m}^{i}_k f_\chi \left(\frac{x - \bar{c}^i_k}{\bar{m}^{i}_k} ;1,\lambda^i_k \right) + \bar{c}^i_k F_\chi \left(\frac{x - \bar{c}^i_k}{\bar{m}^{i}_k} ;1,\lambda^i_k\right) \right]p^{i}_k,
\end{align*}
where $F_\chi(\cdot, k, \lambda)$ and $f_\chi(\cdot, k, \lambda)$ are the distribution and density functions of a non-central chi-square random variable with degrees of freedom $k$ and non-centrality parameter $\lambda$, using the shorthand notation $\bar{m}^{i}_k\coloneqq\bar{m}(x^i_k)$, $\bar{c}^i_k\coloneqq\bar{c}(x^i_k)$ and $\lambda^i_k\coloneqq\bar{\lambda}(x^i_k)$.

\clearpage
\bibliographystyle{apalike}
\bibliography{References}

\end{document}